\pdfoutput=1
\documentclass[10pt,a4paper,english,journal]{IEEEtran}
\usepackage[T1]{fontenc}
\usepackage[latin9]{inputenc}
\usepackage{array}
\usepackage{multirow}
\usepackage{amsmath}
\usepackage{amssymb}
\usepackage{stackrel}
\usepackage{graphicx}

\makeatletter

\pdfpageheight\paperheight
\pdfpagewidth\paperwidth

\providecommand{\tabularnewline}{\\}

\usepackage{xcolor,soul}
\sethlcolor{lightgray}

\makeatother

\usepackage{babel}
\begin{document}

\title{On the Ultra-Reliable and Low-Latency Communications in Flexible
TDD/FDD 5G Networks}

\author{\IEEEauthorblockN{Ali A. Esswie$^{1,2}$,\textit{ Member, IEEE}, and\textit{ }Klaus
I. Pedersen\textit{$^{1,2}$, Senior Member, IEEE}\\
$^{1}$Nokia Bell-Labs, Aalborg, Denmark\\
$^{2}$Department of Electronic Systems, Aalborg University, Denmark}}

\maketitle
$\pagenumbering{gobble}$
\begin{abstract}
The ultra-reliable and low-latency communication (URLLC) is the key
driver of the current 5G new radio standardization. URLLC encompasses
sporadic and small-payload transmissions that should be delivered
within extremely tight radio latency and reliability bounds, i.e.,
a radio latency of 1 ms with $99.999\%$ success probability. However,
such URLLC targets are further challenging in the 5G dynamic time
division duplexing (TDD) systems, due to the switching between the
uplink and downlink transmission opportunities and the additional
inter-cell cross-link interference (CLI). This paper presents a system
level analysis of the URLLC outage performance within the 5G new radio
flexible TDD systems. Specifically, we study the feasibility of the
URLLC outage targets compared to the case with the 5G frequency division
duplexing (FDD), and with numerous 5G design variants. The presented
results therefore offer valuable observations on the URLLC outage
performance in such deployments, and hence, introducing the state-of-the-art
flexible-FDD technology. 

\end{abstract}

\section{Introduction}

The fifth generation (5G) new radio (NR) is designed to support a
variety of services such as ultra-reliable and low-latency communications
(URLLC) {[}1{]}, industrial time sensitive communications (TSC) {[}2{]},
and enhanced mobile broadband (eMBB) communications {[}3{]}. Those
come with challenging requirements for the packet latency, jitter,
and aggregated capacity, respectively. On another side, dynamic time
division duplexing (TDD) is the major duplexing technology for 5G
NR due to the wide spectrum availability of unpaired bands, i.e.,
the 3.5 GHz band, and spectrum above 6 GHz {[}4{]}. Additionally,
the frequency division duplexing (FDD) is also supported for 5G NR,
and considered especially relevant for deployments at bands below
6 GHz {[}5{]}. In this regard, fulfilling the URLLC requirements for
FDD systems is obviously more manageable since both base-stations
(BSs) and user-equipments (UEs) always have simultaneous uplink (UL)
and downlink (DL) transmission opportunities. 

Although, for TDD deployments, it is further challenging to fulfill
such targets due to the restriction of either having exclusively UL
or DL transmissions. Hence, in a multi-cell multi-user scenario, it
becomes a hard problem to ensure that the URLLC latency and reliability
requirements are met for all active UEs, as the inter-UE timing relations
may likely be different. It is therefore a non-trivial problem how
to dynamically adjust the UL-DL switching for 5G NR TDD. 

The standardization body has accordingly defined a flexible slot format
design {[}6{]}, where the traffic adaptation could occur per 14-OFDM
symbol slots. In principle, such a design allows BSs to dynamically
adapt their link directions, i.e., UL or DL symbols, according to
a local selection criterion such as the buffered traffic statistics
(incl. e.g., the related head of line delay). Although, when different
neighboring BSs concurrently adopt opposite transmission link directions,
it comes with the cost of potentially severe cross-link interference
(CLI) {[}7{]}. CLI is highly critical for achieving the URLLC outage
requirements, where especially the BS-BS CLI is problematic due to
the higher BS transmit power as compared to the UE transmit powers.
Accordingly, the majority of the recent TDD studies tackled the CLI
issue either by pre-avoidance or post-cancellation techniques. In
{[}8{]}, coordinated inter-cell user scheduling, and advanced UL power
control are introduced to minimize the average network CLI. Furthermore,
opportunistic frame coordination schemes {[}7, 9{]} are proposed to
pre-avoid the occurrence of the BS-BS and UE-UE CLI on a best-effort
basis. Moreover, perfect BS-BS CLI cancellation using full packet
exchange and orthogonal projector estimation are discussed in {[}10,
11{]}. 

In this paper, we study the URLLC outage performance in an advanced
system-level setting with high degree of realism. Particularly, how
to most efficiently manage the switching between the UL and DL transmission
opportunities to best meet the URLLC traffic conditions is investigated,
assuming bi-directional random time-variant traffic. The impact of
adjusting the TDD switching pattern at different time-resolutions
is analyzed, including a sensitivity analysis for other system-level
parameter settings and algorithm variants. For dynamic TDD, we isolate
the effect of the CLI by presenting both cases where the CLI is realistically
modeled, in addition to the case where an optimal CLI cancellation
is assumed. To the best of our knowledge, no prior studies have presented
such system-level URLLC outage results and related recommendations
for 5G NR TDD deployments.

This paper is organized as follows. Section II discusses the system
modeling. Section III introduces the URLLC radio latency analysis
in dynamic-TDD systems, while Section IV presents our adaptation criterion
of the dynamic link selection. The URLLC outage latency assessment
is introduced in Section V. Finally, the flexible-FDD duplexing mode
is discussed in Section VI, while conclusions appear in Section VII. 

\section{System Modeling }

We consider a 5G-NR dynamic TDD macro network with $C$ BSs, each
with $N_{t}$ antennas, where there are $K^{\textnormal{dl}}$ and
$K^{\textnormal{ul}}$ uniformly-distributed DL and UL active UEs
per BS, each with $M_{r}$ antennas. We assume inter-BS synchronized
TDD transmissions, as depicted in Fig. 1. Additionally, the URLLC-alike
sporadic FTP3 traffic model is adopted with the packet sizes of $\textnormal{\ensuremath{\mathit{f}^{dl}}}$
and $\textnormal{\ensuremath{\mathit{f}^{ul}}}$ bits, and Poisson
Point Processes, with mean packet arrivals $\textnormal{\ensuremath{\lambda}}^{\textnormal{dl}}$
and $\textnormal{\ensuremath{\lambda}}^{\textnormal{ul}},$ in the
DL and UL directions, respectively. Thus, the average offered traffic
load per BS in the DL direction is expressed as: $\varOmega^{\textnormal{dl}}=$$K^{\textnormal{dl}}\times\textnormal{\ensuremath{\mathit{f}^{dl}}}\times\textnormal{\ensuremath{\lambda}}^{\textnormal{dl}}$,
and in the UL direction as: $\varOmega^{\textnormal{ul}}=K^{\textnormal{ul}}\times\textnormal{\ensuremath{\mathit{f}^{ul}}}\times\textnormal{\ensuremath{\lambda}}^{\textnormal{ul}}$.
The total offered load per BS is: $\varOmega=\varOmega^{\textnormal{dl}}+\varOmega^{\textnormal{ul}}.$

\begin{figure}
\begin{centering}
\includegraphics[scale=0.7]{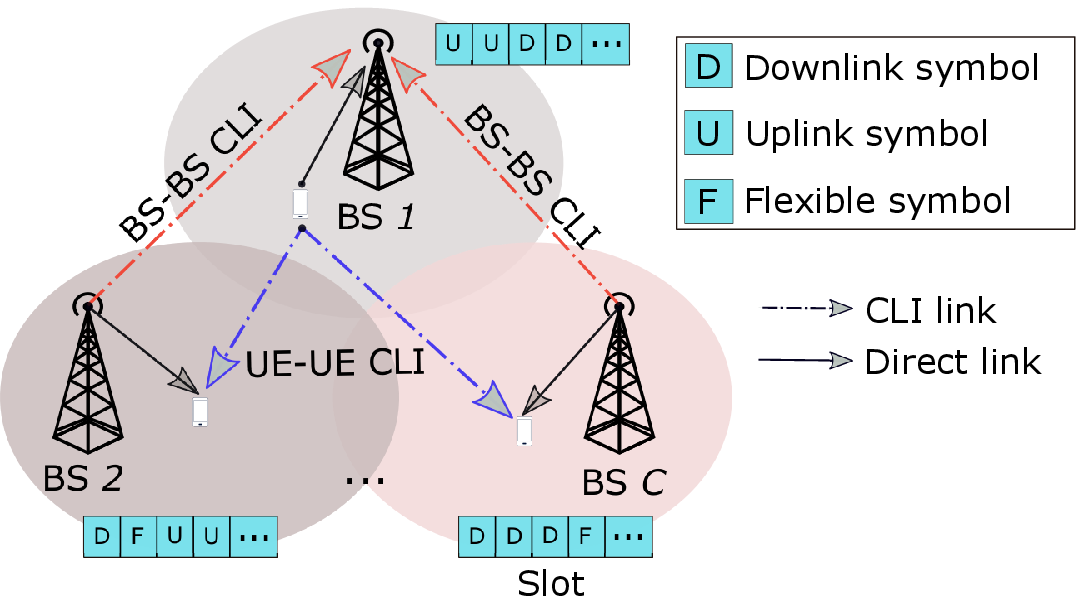}
\par\end{centering}
\centering{}{\small{}Fig. 1. Dynamic-TDD deployment per slot periodicity.}{\small \par}
\end{figure}

We adopt the state-of-the-art 3GPP 5G-NR configurations. UEs are multiplexed
using the orthogonal frequency division multiple access (OFDMA), and
with 30 kHz sub-carrier spacing (SCS). The smallest resource unit,
granted to an active UE, is the physical resource block (PRB) of 12
consecutive SCs. The dynamic user scheduling is applied per a TTI
duration of 4-OFDM symbols, for faster URLLC transmissions. 

\section{URLLC Radio Latency Analysis }

The 3GPP 5G-NR release-15 standard has defined several slot format
designs {[}6{]}. A slot format denotes a certain placement of the
DL {[}D{]}, UL {[}U{]}, and flexible {[}F{]}, OFDM symbols within
a slot duration of 14 OFDM symbols. The flexible symbols imply that
these could be used either for UL/DL transmissions or as guard intervals
between consecutive DL and UL symbols. The average one-way URLLC latency
in the DL direction $\Psi_{\textnormal{dl}}$ is given by 

\begin{equation}
\Psi_{\textnormal{dl}}=\varLambda_{\textnormal{bsp}}+\psi_{\textnormal{tq}}+\psi_{\textnormal{fa}}+\psi_{\textnormal{tti}}+\alpha\psi_{\textnormal{harq}}+\varLambda_{\textnormal{uep}},
\end{equation}
where $\varLambda_{\textnormal{bsp}},$ $\psi_{\textnormal{tq}},$
$\psi_{\textnormal{fa}},$ $\psi_{\textnormal{tti}},$ $\psi_{\textnormal{harq}}$
and $\varLambda_{\textnormal{uep}}$ denote the BS processing, DL
total queuing, DL frame alignment, DL packet transmission, DL hybrid
automatic repeat request (HARQ) re-transmission, and UE processing
delays, respectively. $\alpha$ implies the target block error rate
(BLER), e.g., for a URLLC-alike BLER = 1\%, $\alpha=0.01,$ and $\alpha=0$
if the packet has been successfully decoded from the first transmission.
As can be observed, $\varLambda_{\textnormal{bsp}},$ $\psi_{\textnormal{tti}}$
and $\varLambda_{\textnormal{uep}}$ impose a constant delay offset,
and are only dependent on the UE/BS processing capabilities, and TTI
size, respectively; however, $\psi_{\textnormal{tq}}$ and $\psi_{\textnormal{harq}}$
are time-varying DL delay components, depending on the DL offered
load level, DL and UL link switching delay, and the inflicted DL interference,
respectively. 

Accordingly, the DL HARQ delay $\psi_{\textnormal{harq}}$ is expressed
as

\begin{equation}
\psi_{\textnormal{harq}}=\varLambda_{\textnormal{uep}}+\varphi_{\textnormal{fa}}+\varphi_{\textnormal{nack}}+\varLambda_{\textnormal{bsp}}+\psi_{\textnormal{tq}}+\psi_{\textnormal{fa}}+\psi_{\textnormal{tti}},
\end{equation}
where $\varphi_{\textnormal{fa}}$ implies the alignment delay towards
the first UL control channel opportunity for the UE to transmit the
HARQ negative acknowledgment (NACK), with $\varphi_{\textnormal{nack}}$
as the NACK transmission time. The summation $\varphi_{\textnormal{fa}}+\varphi_{\textnormal{nack}}+\varLambda_{\textnormal{bsp}}$
represents the total delay from the time a UE has identified a corrupted
DL packet until the BS becomes aware of it. Subsequently, the total
DL queuing delay $\psi_{\textnormal{tq}}$ is calculated by

\begin{equation}
\psi_{\textnormal{tq}}=\psi_{\textnormal{q}}+\psi_{\textnormal{tdd}},
\end{equation}
where $\psi_{\textnormal{q}}$ implies the packet queuing delay due
to the dynamic multi-user scheduling, and $\psi_{\textnormal{tdd}}$
is TDD UL-DL link-switching delay, i.e., the additional DL buffering
delay towards the first available DL transmission symbol(s) due to
the non-concurrent DL and UL transmission availability. For instance,
with FDD, $\psi_{\textnormal{tdd}}=0$ ms. Fig. 2a shows an example
of the factors which contribute to the average one-way DL latency
$\Psi_{\textnormal{dl}}$, where a single DL packet associated with
one HARQ re-transmission is assumed. As can be observed, the DL packet
is decoded at its intended UE after 22 OFDM-symbol duration, i.e.,
$0.7$ ms, from its arrival time at the BS, satisfying the URLLC 1-ms
radio latency target; however, with the assumption of immediate DL
scheduling and transmission once the packet arrives the BS DL buffer,
i.e., $\psi_{\textnormal{td}}=0.$ 
\begin{figure*}
\begin{centering}
\includegraphics[scale=0.7]{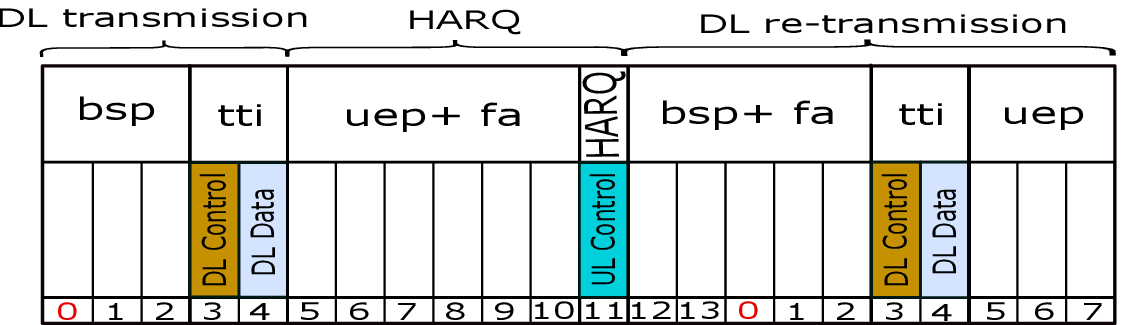}
\par\end{centering}
\centering{}{\small{}(a) Latency of a single DL packet arrival.}{\small \par}
\end{figure*}

Similarly, the one-way URLLC UL latency $\Psi_{\textnormal{ul}}$
follows a similar behavior as $\Psi_{\textnormal{dl}}$; however,
with a linear delay offset due to the UL scheduling. Specifically,
with dynamic-grant (DG) UL scheduling, UEs first align to the first
available transmission opportunity of the UL control channel, i.e.,
$\varphi_{\textnormal{fa}}$, in order to send the scheduling request
(SR), and accordingly wait for the scheduling grant (SG) from the
serving BS over the DL control channel. Thus, $\Psi_{\textnormal{ul}}$
is given by 
\begin{figure*}
\begin{centering}
\includegraphics[scale=0.7]{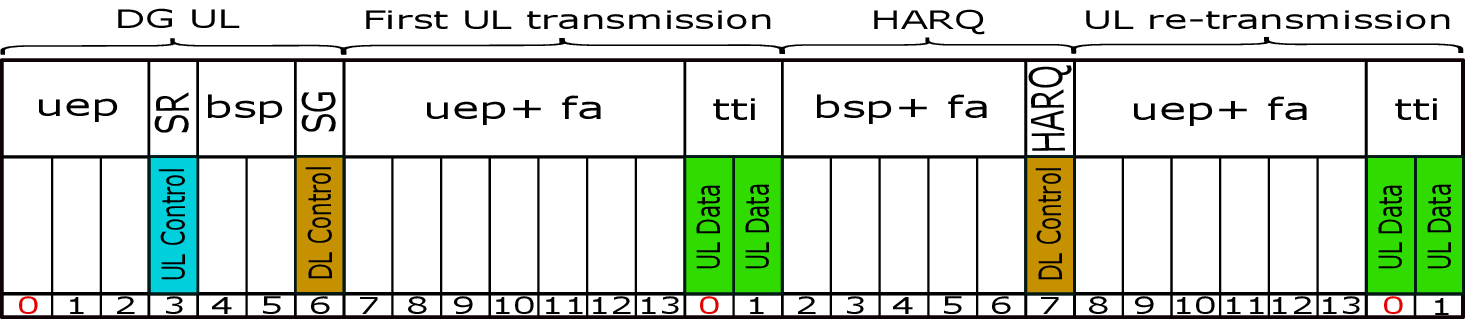}
\par\end{centering}
\centering{}{\small{}(b) Latency of a single UL packet arrival, with
DG UL scheduling.}{\small \par}
\end{figure*}
\begin{figure*}
\begin{centering}
\includegraphics[scale=0.6]{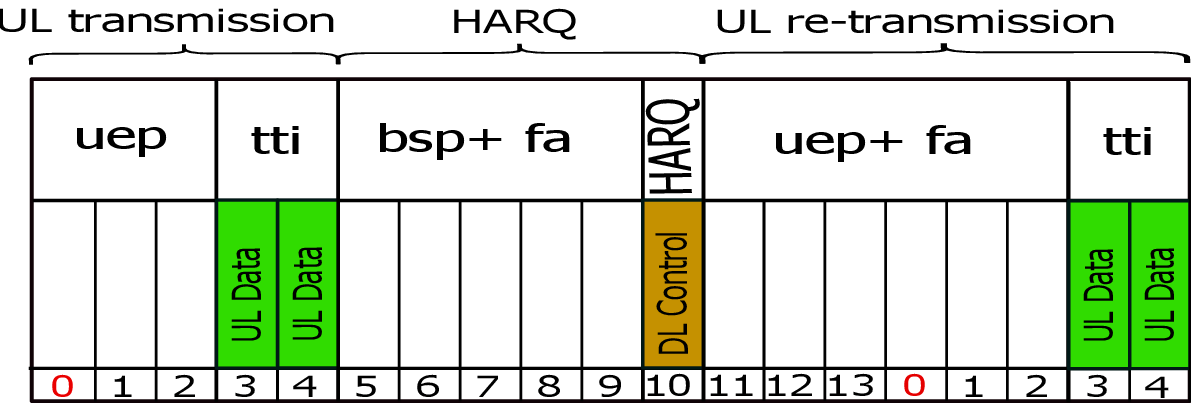}
\par\end{centering}
\centering{}{\small{}(c) Latency of a single UL packet arrival, with
GF UL scheduling.}{\small \par}
\end{figure*}

\begin{equation}
\Psi_{\textnormal{ul}}=\varphi_{\textnormal{dg}}+\varphi_{\textnormal{td}}+\varphi_{\textnormal{fa}}+\varphi_{\textnormal{tti}}+\alpha\varphi_{\textnormal{harq}}+\varLambda_{\textnormal{bsp}},
\end{equation}
where $\varphi_{\textnormal{dg}},$ $\varphi_{\textnormal{td}},$
$\varphi_{\textnormal{fa}},$ $\varphi_{\textnormal{tti}}$ and $\varphi_{\textnormal{harq}}$
are the UL DG delay, UL total buffering delay, UL frame alignment
delay, UL payload transmission delay, and UL HARQ delay, respectively.
\begin{figure*}
\begin{centering}
\includegraphics[scale=0.6]{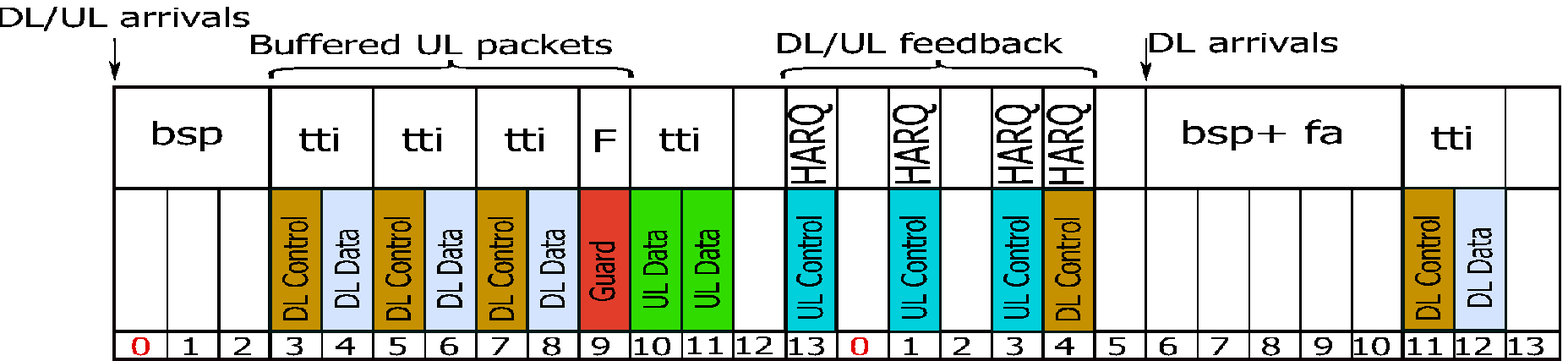}
\par\end{centering}
\begin{centering}
{\small{}(d) Latency of multiple UL and DL packet arrivals.}
\par\end{centering}{\small \par}
\centering{}Fig. 2. URLLC one-way latency components with DG, and
GF UL, for a TTI size = 2-OFDM symbols, and SCS = 30 kHz.
\end{figure*}

On another side, the grant-free (GF) UL scheduling {[}12{]} is considered
as vital for URLLC UL transmissions. With UL grant-free, sporadic
UL packets become immediately eligible for scheduling and transmission,
i.e., no SR and SG delays, $\varphi_{\textnormal{dg}}=\varphi_{\textnormal{sr}}=\psi_{\textnormal{sg}}=0$
ms, with $\varphi_{\textnormal{sr}}$ and $\psi_{\textnormal{sg}}$
as the transmission delays of the SR and SG, respectively; although,
with the DG, $\varphi_{\textnormal{dg}}$ is then calculated as 

\begin{equation}
\varphi_{\textnormal{dg}}=\varLambda_{\textnormal{uep}}+\varphi_{\textnormal{fa}}+\varphi_{\textnormal{sr}}+\varLambda_{\textnormal{bsp}}+\psi_{\textnormal{fa}}+\psi_{\textnormal{sg}}+\varLambda_{\textnormal{uep}}^{\text{'}},
\end{equation}
with $\varLambda_{\textnormal{uep}}^{\text{'}}$ as the UE processing
delay to decode the SG, i.e., $\varLambda_{\textnormal{uep}},$ as
well as preparing the UL transport block, where $\varLambda_{\textnormal{uep}}^{\text{'}}>\varLambda_{\textnormal{uep}}.$

Equivalently to (2) and (3), the UL HARQ $\varphi_{\textnormal{harq}}$
and total queuing $\varphi_{\textnormal{td}}$ delays are given by

\begin{equation}
\varphi_{\textnormal{harq}}=\varLambda_{\textnormal{bsp}}+\psi_{\textnormal{fa}}+\psi_{\textnormal{nack}}+\varLambda_{\textnormal{uep}}+\varphi_{\textnormal{td}}+\varphi_{\textnormal{fa}}+\varphi_{\textnormal{tti}}.
\end{equation}

\begin{equation}
\varphi_{\textnormal{td}}=\varphi_{\textnormal{q}}+\varphi_{\textnormal{tdd}},
\end{equation}
where $\varphi_{\textnormal{q}}$ and $\varphi_{\textnormal{tdd}}$
denote the UL packet queuing delay and the delay towards the first
available UL transmission opportunity, where $\varphi_{\textnormal{td}}\neq\psi_{\textnormal{td}}$
due to the different UL and DL offered load, leading to varying UL
and DL buffering performance, respectively. Fig. 2b and 2c depict
the radio latency components which affect the average one-way URLLC
UL latency $\Psi_{\textnormal{ul}},$ for the DG and GF UL scheduling
cases, respectively, and under the assumption of a single UL packet
arrival without further multi-UE queuing delays. With the UL DG and
one UL HARQ re-transmission, the URLLC UL packet gets delivered after
30-OFDM symbol duration, i.e., $1$ ms, which does not allow for any
further packet buffering due to the dynamic user scheduling; otherwise,
the URLLC UL 1-ms latency target shall be violated. 

Finally, Fig. 2d presents an example of multiple concurrent DL and
UL packet arrivals, unlike Fig. 2a, 2b, and 2c, respectively. Herein,
the BS decides multiple DL TTIs first to transmit the early-arriving
DL packets. Accordingly, the UL packets are buffered over those DL
TTIs as well as several guard symbols towards the first available
UL TTI opportunity, i.e., $\varphi_{\textnormal{tdd}}\ggg0$ ms, exceeding
the UL latency budget. Next, the BS adopts alternating DL and UL TTI
instances for the subsequent DL/UL HARQ feedback. 

\section{Traffic adaptation in dynamic-TDD systems}

For a dynamic TDD deployment, BSs dynamically match their transmission
link directions to the sporadic traffic arrivals. Hence, at each pattern
update periodicity, which could be either per a slot or aggregated
several slots, BSs select the slot formats, i.e., number of DL and
UL symbols during the next slot(s), which best satisfy their individual
link direction selection criteria. We consider the amount of buffered
DL and UL traffic to select the link directions. Thus, we define the
buffered traffic ratio $\varpi_{c}$ as 

\begin{equation}
\varpi_{c}=\frac{Z_{c}^{\textnormal{dl}}}{Z_{c}^{\textnormal{dl}}+Z_{c}^{\textnormal{ul}}},
\end{equation}
where $Z_{c}^{\textnormal{dl}}$ and $Z_{c}^{\textnormal{ul}}$ are
the aggregated buffered traffic size in the DL and UL directions,
respectively. Herein, we assume perfect knowledge of $Z_{c}^{\textnormal{ul}}$
at the BSs from the UEs buffer status reports and pending SRs, respectively.
The lower $\varpi_{c}$ ratio, the larger the buffered UL traffic
volume, and thus, BSs select slot formats with a majority of UL symbols.
For instance, at an arbitrary BS with $\varpi_{c}=0.2$, the buffered
UL traffic volume is $\textnormal{\ensuremath{4\textnormal{x}}}$
the buffered DL traffic, thus, BS consequently selects a slot format
of DL:UL symbol ratio as $\sim1:4$. In case there are neither new
packet arrivals nor buffered traffic at an arbitrary time instant,
BSs fall back to a default slot format with equal DL and UL symbol
share; however, BSs do not schedule any UEs though. This way, BSs
tend to rapidly adapt to the accumulating buffered traffic, equalizing
both the DL and UL TDD queuing performance, i.e., $\psi_{\textnormal{tdd}}$
and $\varphi_{\textnormal{tdd}}$. 

In this work, the order of the DL and UL OFDM symbols during the adopted
slot format(s) is evenly distributed with a block size of 4 symbols,
e.g., a selected slot pattern of $\sim2:1$ DL:UL symbol ratio is
configured as: {[}DDDDFUUUUDDDDF{]}. Such configuration allows for
alternating DL and UL transmission opportunities during each slot
duration for urgent packet arrivals; however, it comes at the expense
of inflicting more guard symbols, i.e., {[}F{]} symbols, among each
DL and UL symbol pair. 

\section{URLLC Outage Latency Assessment }

We evaluate the URLLC radio performance using inclusive system level
simulations {[}7{]}, where the major functionalities of the physical
and media access control layers, respectively, are implemented according
to the latest 5G-NR specifications. The default simulation assumptions
are listed in Table I, unless otherwise mentioned. We consider asynchronous
Chase-combining HARQ, where the HARQ re-transmissions are dynamically
scheduled and always prioritized over new transmissions. Finally,
the URLLC outage latency, i.e., radio latency at the $10^{-5}$ outage
probability, is assessed under various 5G system configurations. 

\begin{table}
\caption{{\small{}Default simulation parameters.}}
\centering{}%
\begin{tabular}{c|c}
\hline 
Parameter & Value\tabularnewline
\hline 
Environment & 3GPP-UMA, one cluster, 21 cells\tabularnewline
\hline 
UL/DL channel bandwidth & 20 MHz, SCS = 30 KHz, TDD\tabularnewline
\hline 
Antenna setup & $N_{t}=4$, $M_{r}=4$\tabularnewline
\hline 
UL power control & $\alpha=1,\:P0=-103$ dBm\tabularnewline
\hline 
Link adaptation & Adaptive modulation and coding \tabularnewline
\hline 
UE processing time & $\begin{array}{c}
\textnormal{DL : 4.5/9-OFDM symbols}\\
\textnormal{UL : 5.5/11-OFDM symbols}
\end{array}$\tabularnewline
\hline 
Average user load per cell & $K^{\textnormal{dl}}=K^{\textnormal{ul}}=$ 1, 10, 50, 100 and 200 \tabularnewline
\hline 
TTI configuration & 4-OFDM symbols\tabularnewline
\hline 
Traffic model & $\begin{array}{c}
\textnormal{\textnormal{FTP3}}\\
\textnormal{\textnormal{\ensuremath{\mathit{f}^{dl}}} = \textnormal{\ensuremath{\mathit{f}^{ul}}} = 400 bits}\\
\textnormal{\textnormal{\ensuremath{\textnormal{\ensuremath{\lambda}}^{\textnormal{ul}}} = \textnormal{\ensuremath{\textnormal{\ensuremath{\lambda}}^{\textnormal{dl}}} = 100 pkts/sec}}}
\end{array}$\tabularnewline
\hline 
Interference conditions & Interference-free\tabularnewline
\hline 
DL/UL scheduling & Proportional fair; UL GF {[}12{]}\tabularnewline
\hline 
DL/UL receiver & LMMSE-IRC\tabularnewline
\hline 
Pattern update periodicity & 1 radio frame (10 ms)\tabularnewline
\hline 
\end{tabular}
\end{table}

\textbf{\textit{URLLC outage latency with pattern update periodicity
$\gamma$:}}

The pattern update periodicity implies how frequent the BSs update
their corresponding slot formats, hence, how fast they adapt the network
capacity towards the sporadic DL/UL packet arrivals. For instance,
\textbf{\textit{$\gamma=1$}} slot denotes that BSs update their adopted
DL and UL symbol patterns per every slot duration, i.e., 14 OFDM symbols.
Fig. 3 holds a comparison of the DL/UL combined URLLC outage latency
under FDD, TDD with \textbf{\textit{$\gamma=1$}} slot and a single
frame duration, respectively, and for 20 MHz bandwidth. An equivalent
FDD bandwidth allocation is also adopted, i.e., 10 MHz for UL transmissions
and 10 MHz for DL transmissions. As can be clearly seen, at the lower
load, i.e., $\varOmega=0.5$ Mbps, both duplexing schemes under evaluation
achieve the 1-ms URLLC latency target. Although, by increasing the
offered load up to $\varOmega=2.5$ Mbps, the FDD significantly outperforms
the respective TDD, in terms of the URLLC outage latency due to the
immediate availability of the DL and UL capacity, i.e., no TDD delays
inflicted, and hence, $\psi_{\textnormal{tdd}}=\varphi_{\textnormal{tdd}}=0$
ms. Accordingly, the TDD with a \textbf{\textit{$\gamma=1$}} slot
achieves a greatly improved URLLC outage latency, i.e., $-218.7\%$
outage latency reduction compared to the case of \textbf{\textit{$\gamma=1$}}
radio-frame, because of the faster link adaptation to the random DL/UL
packet arrivals, leading to less traffic buffering delays. However,
this comes with a significantly increased control overhead size, due
to the guard time duration between each consecutive DL and UL symbol
pair. 

\begin{figure}
\begin{centering}
\includegraphics[scale=0.6]{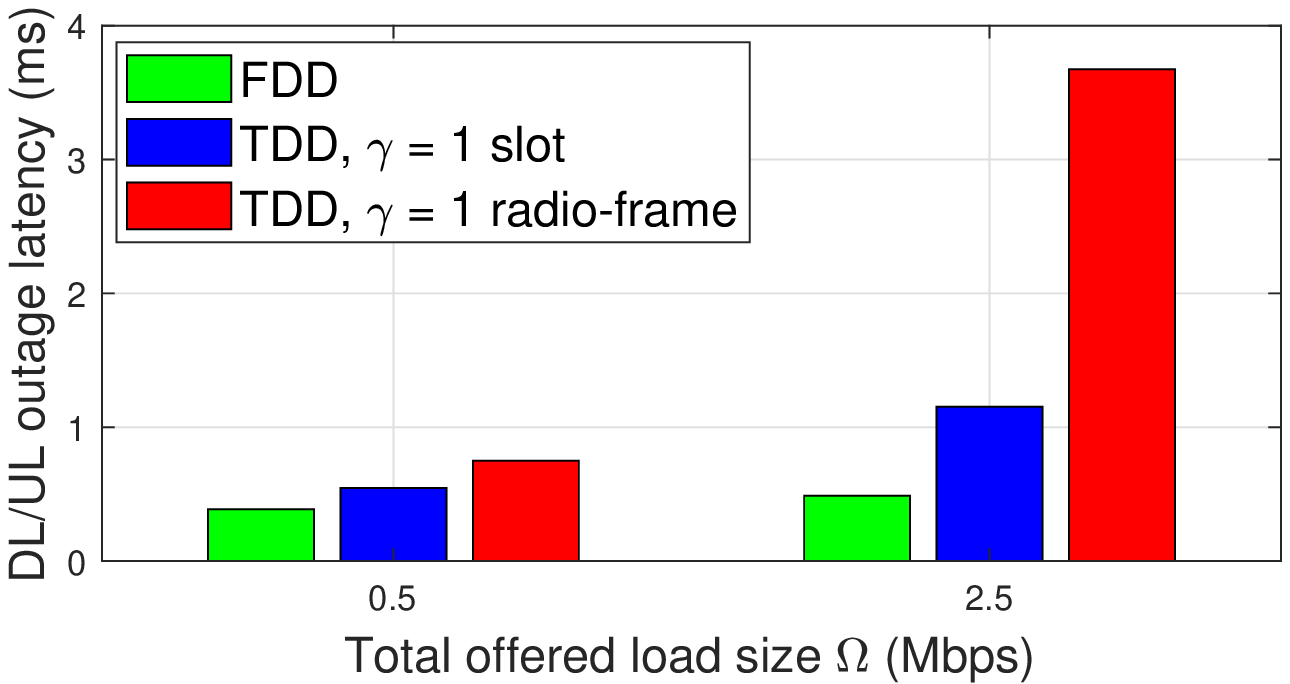}
\par\end{centering}
\centering{}{\small{}Fig. 3. URLLC outage latency: with }\textbf{\textit{$\gamma$}}{\small{}. }{\small \par}
\end{figure}

\textbf{\textit{URLLC outage latency with dynamic and grant-free UL:}}

Based on the latency analysis in Section III, grant-free UL has been
demonstrated to significantly reduce the URLLC UL outage latency,
compared to DG. Accordingly, Fig. 4 depicts the complementary cumulative
distribution function (CCDF) of the URLLC DL/UL combined latency when
UL grant-free and DG are adopted. Herein, with DG UL, UEs  transmit
the scheduling request on a periodicity of 16 TTIs, and hence, receive
the corresponding scheduling grant 4 TTIs later. As noticed, the DG
UL exhibits a linear offset in the UL outage latency by the additional
latency component $\varphi_{\textnormal{dg}}$, leading to $\sim+400\%$
increase in the URLLC outage latency, compared to the GF UL case,
with $\varphi_{\textnormal{dg}}=0$ ms.

\begin{figure}
\begin{centering}
\includegraphics[scale=0.6]{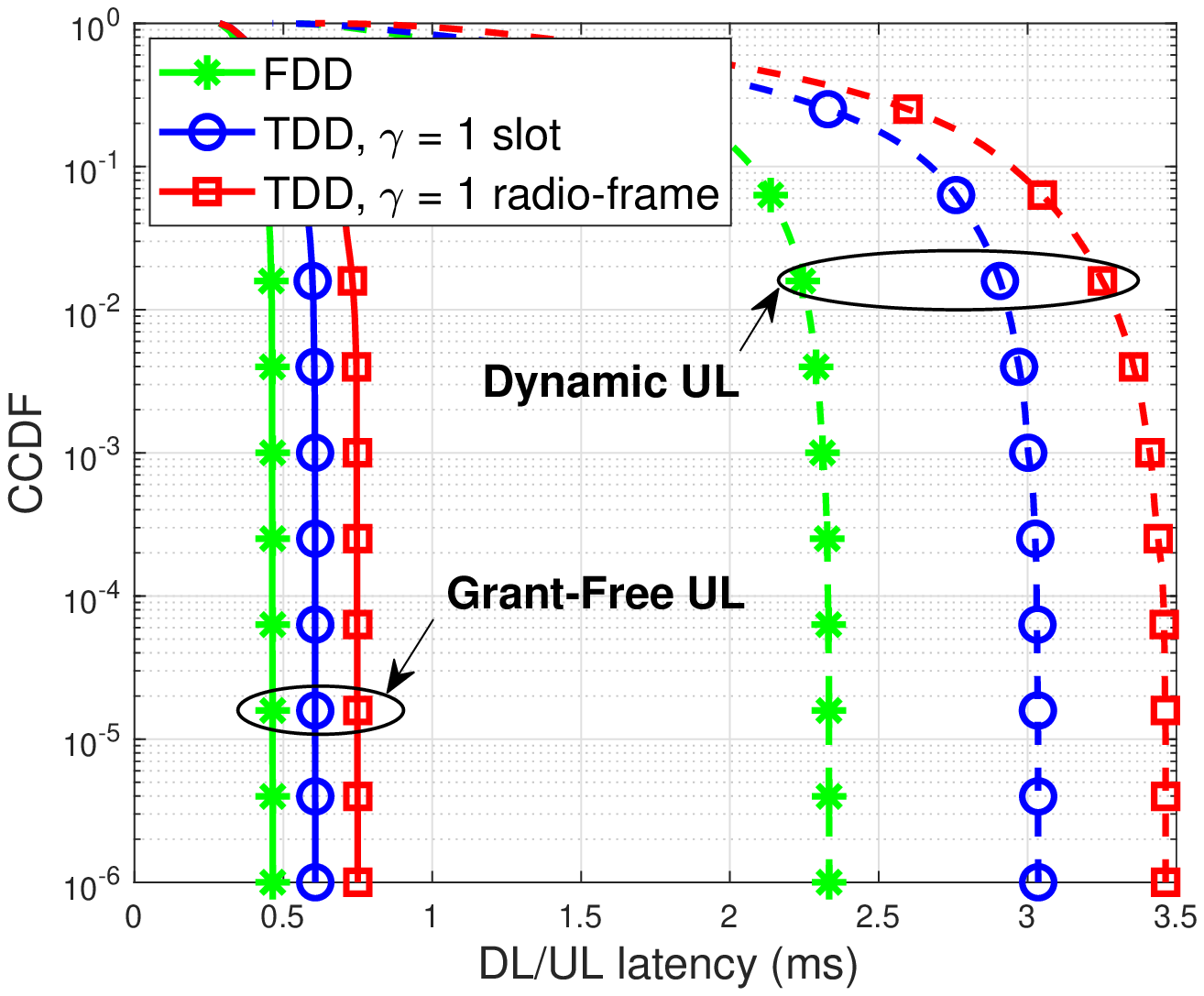}
\par\end{centering}
\centering{}{\small{}Fig. 4. URLLC outage latency: with DG, GF, }$\varOmega=0.5${\small{}
Mbps. }{\small \par}
\end{figure}
 
\begin{figure}
\begin{centering}
\includegraphics[scale=0.5]{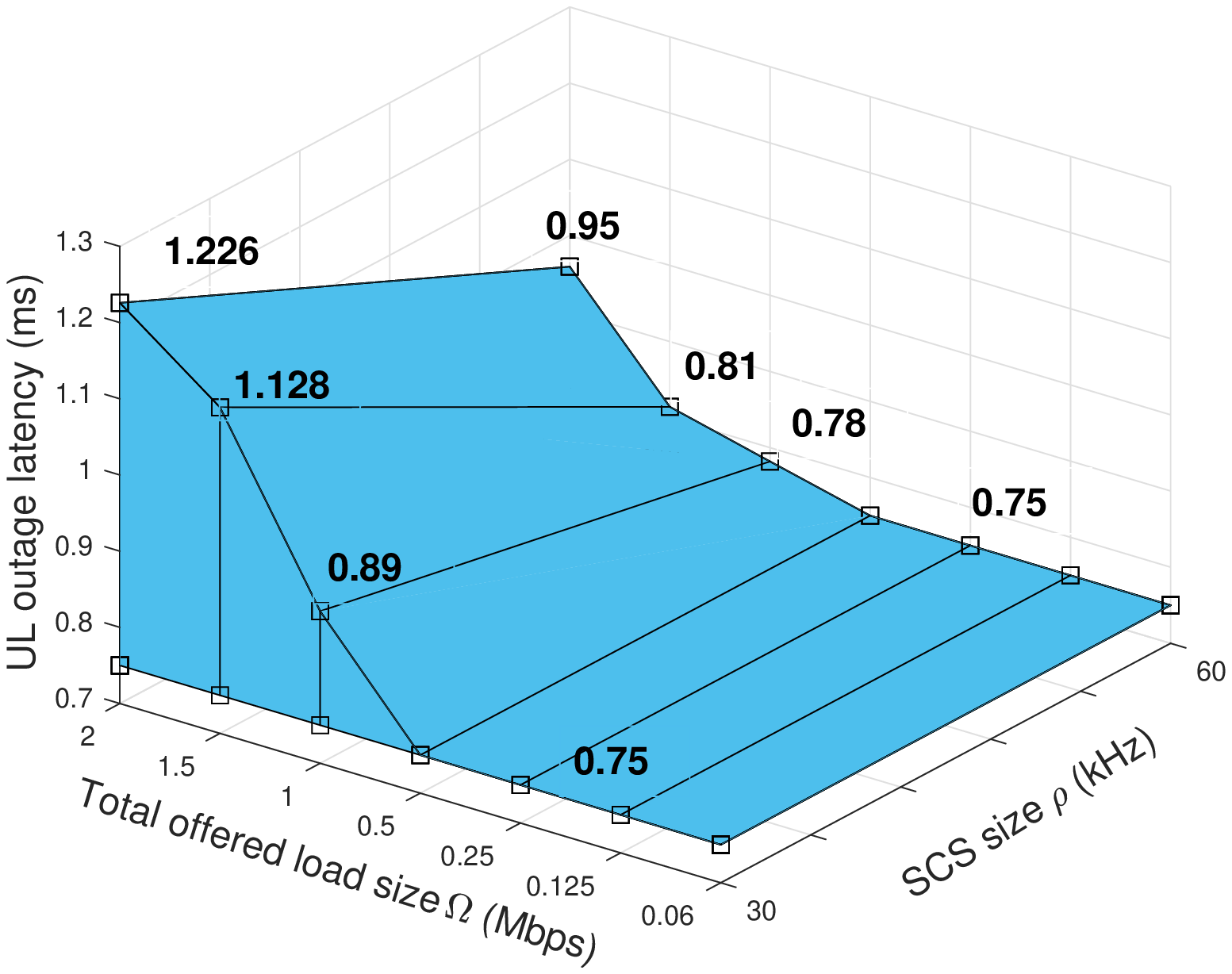}
\par\end{centering}
\centering{}{\small{}Fig. 5. URLLC UL outage latency: with }\textbf{\textit{$\rho$}}{\small{}. }{\small \par}
\end{figure}

\textbf{\textit{URLLC outage latency with the SCS size $\rho$: }}

The size of the channel SCS has a critical impact on the URLLC outage
latency. Unlike the 4G standards, the 5G-NR specs adopt different
SCSs for its diverse service classes, i.e., \textbf{\textit{$\rho=15,\,30,$}}
and $60$ kHz, respectively, for the carrier frequencies below 6 GHz.
However, it was recently agreed within the 3GPP community that \textbf{\textit{$\rho=15$}}
kHz is no longer appropriate for URLLC transmissions. Accordingly,
the achievable URLLC UL outage latency with the SCS size is presented
in Fig. 5, for different offered loads $\varOmega$. The larger SCS
size, i.e., \textbf{\textit{$\rho=60$}} kHz, offers: (a) reduced
BS and UE processing delays, i.e., $\varLambda_{\textnormal{bsp}}$
and $\varLambda_{\textnormal{uep}}$, due to the shorter OFDM symbols
in time, and (2) a higher probability of non-segmented URLLC transmissions,
i.e., URLLC payload is transmitted in a single-shot without segmentation,
reducing the DL $\psi_{\textnormal{q}}$ and UL $\varphi_{\textnormal{q}}$
buffering delays, respectively. Accordingly, a larger \textbf{\textit{$\rho$}}
allows for faster URLLC transmissions to compensate for the additional
DL and UL switching delay of the dynamic-TDD systems, satisfying the
stringent URLLC 1-ms outage latency. 

\textbf{\textit{URLLC outage latency with the TTI size $\mu$:}}

The TTI length determines the packet transmission periodicity. Hence,
it has a key impact on the maximum alignment delay that an arbitrary
packet may inflict until the first available DL/UL TTI instance. As
shown in Fig. 6, the empirical CDF (ECDF) of the average scheduling
delay of the combined DL $\psi_{\textnormal{td}}$ and UL $\varphi_{\textnormal{td}}$
transmissions, is introduced for \textbf{\textit{$\mu=4,\,7,$}} and
$14$ OFDM symbols, respectively. Hence, the scheduling delay defines
the delay between the time instant a packet arrives at the scheduling
buffers until it is being transmitted, excluding the processing times.
Obviously, the larger \textbf{\textit{$\mu$}}, the larger the time
delay of which the incoming packets shall exhibit in the scheduling
buffers. The TDD case with \textbf{\textit{$\gamma=1$}} radio-frame
and \textbf{\textit{$\mu=14$}} OFDM symbols clearly provides the
worst scheduling delay performance because of the slower traffic adaptation
periodicity \textbf{\textit{$\gamma$}} and the large TTI alignment
delay, respectively. However, the FDD mode inflicts a lower scheduling
delay due to the absence of the TDD switching delay, i.e., $\psi_{\textnormal{tdd}}=\varphi_{\textnormal{tdd}},=0$
ms. 

\begin{figure}
\begin{centering}
\includegraphics[scale=0.65]{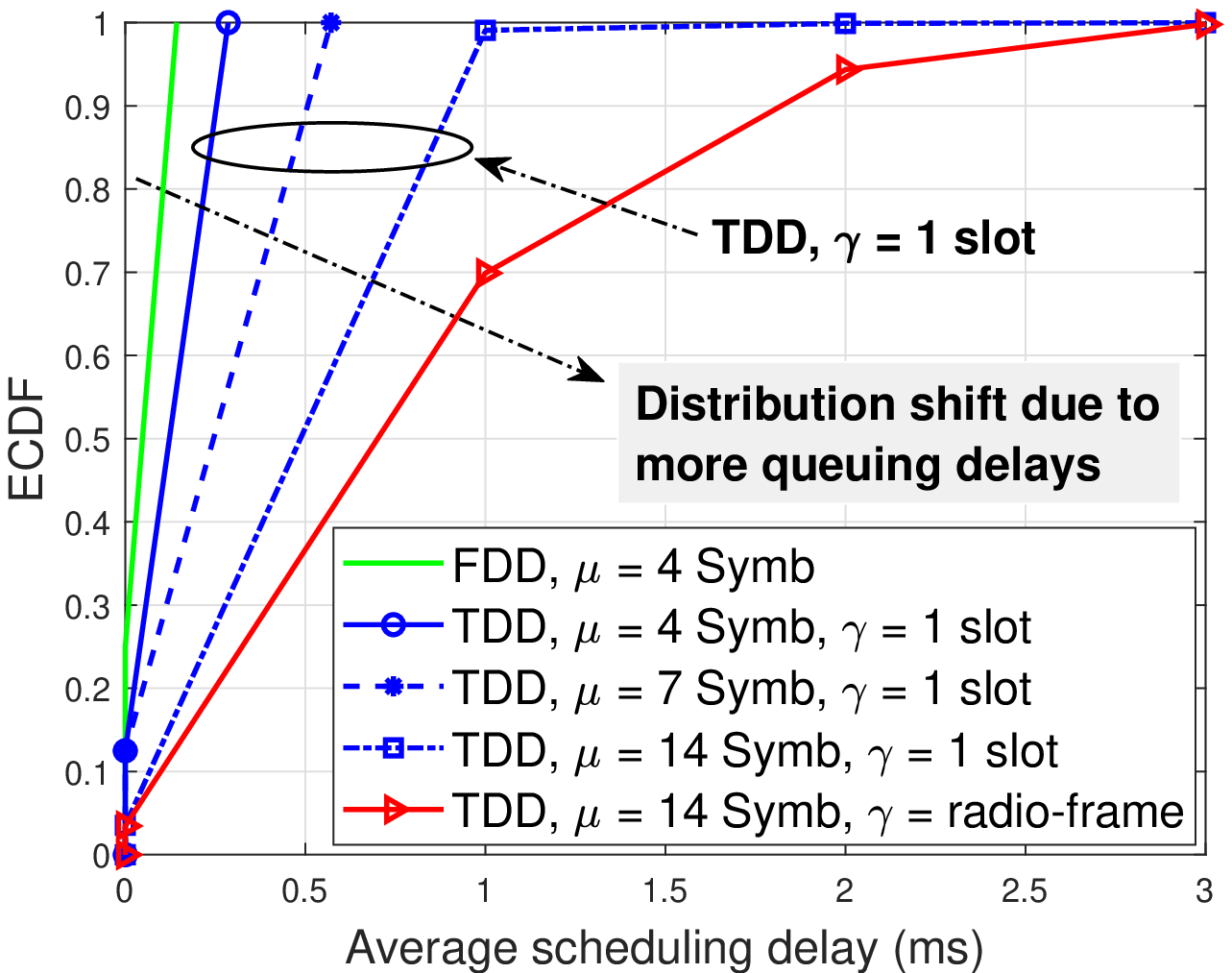}
\par\end{centering}
\centering{}{\small{}Fig. 6. URLLC outage latency: with }\textbf{\textit{$\mu,$}}{\small{}
}$\varOmega=1${\small{} Mbps.}{\small \par}
\end{figure}

\textbf{\textit{URLLC outage latency with the inter-BS CLI:}}

The inter-BS CLI is considered as the most critical challenge against
the 5G-NR dynamic-TDD systems. In this regard, Fig. 7 depicts the
CCDF of the URLLC UL latency with the FDD, and TDD duplexing, under
CLI-non-free and CLI-free conditions, respectively. The latter case
denotes a theoretical baseline, i.e., optimal inter-BS CLI cancellation
is assumed, to which we compare the actual performance of the dynamic-TDD
systems with CLI coexistence. The URLLC outage latency with unhandled
CLI exhibits $+162.19\%$ increase compared to the CLI-free case.
This is mainly because of the UL packets getting re-transmitted several
times prior to a successful decoding, due to the severe BS-BS CLI,
leading to significantly large $\varphi_{\textnormal{td}}$ and $\alpha\varphi_{\textnormal{harq}}$
delays. On another side, the FDD case provides the best UL outage
latency, mainly due to the absolute absence of the inter-BS CLI. 

\section{Discussions on state-of-the-art flexible-FDD}

5G-NR dynamic-TDD systems offer a flexible link direction adaptation
to the sporadic URLLC packet arrivals. However, throughout the paper,
it has been demonstrated an extremely challenging task to achieve
the URLLC outage latency and reliability targets in such systems. 

In order to overcome the latency challenges of the dynamic-TDD operation,
we next briefly consider the option of flexible-FDD operation for
unpaired carriers. With the flexible-FDD, a single unpaired carrier
frequency is utilized such that some PRBs are used for DL transmissions,
while others are concurrently adopted for UL transmissions, as depicted
by Fig. 8. Herein, unlike the dynamic-TDD mode, simultaneous DL and
UL transmissions are allowed, while still dynamically adjusting the
amount of DL and UL frequency resources in line with the BS-specific
link selection criterion. For instance, a BS with a buffered traffic
ratio $\varpi_{c}$ of 3:1, adopts 60\% : 20\% DL-to-UL PRB ratio,
while the remaining frequency resources are flexibly configured as
guard bands. The main advantages of the flexible-FDD over dynamic-TDD
mode are as follows: (a) absence of the DL and UL switching delays,
i.e., $\psi_{\textnormal{tdd}}=\varphi_{\textnormal{tdd}},=0$ ms,
and (b) absence of the inter-BS CLI by simpler frequency coordination
techniques.

However, flexible-FDD requires efficient self-interference mitigation
techniques in practice, in order to cope with the power leakage problem,
resulting from the concurrent DL transmissions and UL receptions over
the same PRB set. Accordingly, the self-interference mitigation operation
is typically implemented as a hybrid process of analog interference
suppression and digital interference cancellation. In that sense,
a possible variant of a flexible-FDD deployment would therefore be
to have BSs operating in the flexible-FDD mode, while connected UEs
operate in half-duplex mode, either having an uplink or downlink link
activation at time. Thereby, each BS shall simultaneously serve different
UEs in opposite/same link directions over partially or fully shared
frequency resources; though, without the need of self-interference
mitigation capabilities at the UE-side.

\begin{figure}
\begin{centering}
\includegraphics[scale=0.55]{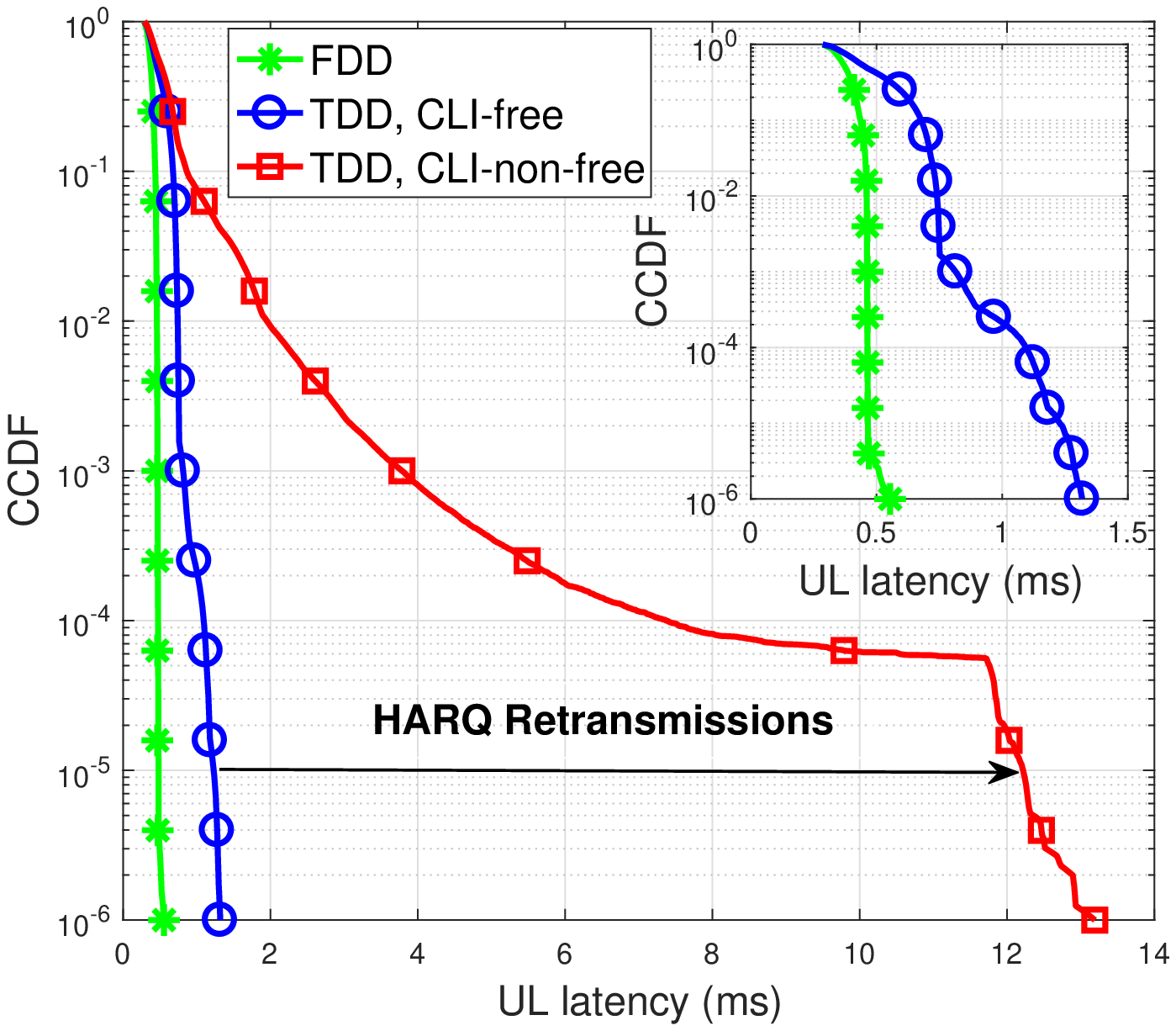}
\par\end{centering}
\centering{}{\small{}Fig. 7. URLLC outage latency: with CLI for }$\varOmega=1.5${\small{}
Mbps. }{\small \par}
\end{figure}

\section{Concluding Remarks }

In this work, we studied the feasibility of the URLLC outage latency
within the 5G new radio dynamic-TDD deployments. The URLLC radio performance
is first evaluated under optimal interference-free conditions, with
the various system design aspects of the 5G new radio, i.e., offered
sporadic packet arrivals, channel sub-carrier spacing, transmission
time interval duration, configured and grant-free uplink scheduling.
Then, the impact of the inter-cell cross link interference on the
achievable URLLC outage latency is identified. Finally, the state-of-the-art
flexible-FDD duplexing mode is being introduced towards the upcoming
3GPP standards. 

The main insights brought by this paper are summarized as follows:
(1) with inter-BS interference-controlled conditions, the 30 kHz sub-carrier
spacing (SCS) is proven suitable to satisfy the URLLC 1-ms radio latency
target for offered loads up to 1 Mbps/BS, (2) with higher offered
load levels, the SCS of 60 kHz and bandwidth allocation of 20 MHz
should be adopted to further reduce the packet segmentation delay,
user scheduling delay, TTI duration, UE and BS processing delays,
(3) dynamic UL scheduling, the BS and UE processing delays, respectively,
introduce a constant delay offset in the URLLC outage latency regardless
of the other system design variants, and hence, they should be particularly
optimized and (4) adding the BS-BS cross-link interference, the URLLC
latency targets are almost not feasible due to the UL capacity blockage.

\begin{figure}
\begin{centering}
\includegraphics[scale=0.55]{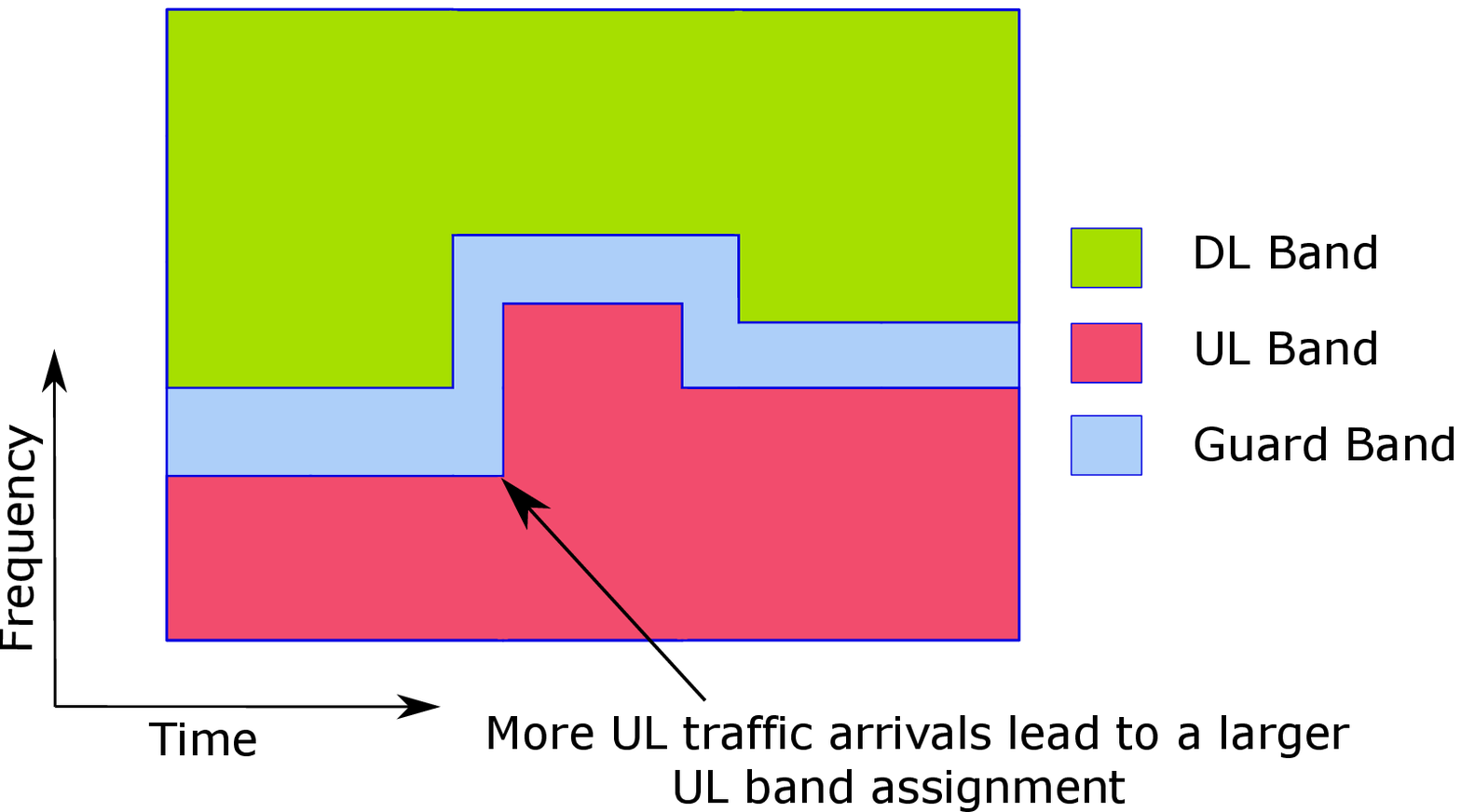}
\par\end{centering}
\centering{}{\small{}Fig. 8. Flexible-FDD towards upcoming 3GPP 5G
standards.}{\small \par}
\end{figure}

\section{Acknowledgments}

This work is partly funded by the Innovation Fund Denmark \textendash{}
File: 7038-00009B.

\end{document}